\documentclass[preprint,aps ,nofootinbib]{revtex4}
\usepackage{graphicx}
\usepackage{amsmath}
\usepackage{amsfonts}
\usepackage{amssymb}
\usepackage{color}%
\usepackage{dcolumn}
\newcommand{\f}{\begin{equation}}
\newcommand{\ff}{\end{equation}}
\newcommand{\fa}{\begin{eqnarray}}
\newcommand{\ffa}{\end{eqnarray}}

\begin{document}
\title{DC and Hall conductivity in holographic massive Einstein-Maxwell-Dilaton gravity}
\author{Zhenhua Zhou $^{1}$}
\email{zhouzh@ihep.ac.cn}
\author{Jian-Pin Wu $^{2,3}$}
\email{jianpinwu@mail.bnu.edu.cn}
\author{Yi Ling $^{1,3}$}
\email{lingy@ihep.ac.cn}

\affiliation{$^1$ Institute of High Energy Physics, Chinese
Academy of Sciences, Beijing 100049, China\\
$^2$ Institute of Gravitation and Cosmology, Department of Physics, School of Mathematics and Physics,
Bohai University, Jinzhou 121013, China \\ $^3$ State Key
Laboratory of Theoretical Physics, Institute of Theoretical
Physics, Chinese Academy of Sciences, Beijing 100190, China}

\begin{abstract}
We investigate the holographic DC and Hall conductivity in massive
Einstein-Maxwell-Dilaton (EMD) gravity. Two special EMD
backgrounds are considered explicitly. One is dyonic
Reissner-Nordstr$\ddot{o}$m-AdS (RN-AdS) geometry and the other
one is hyperscaling violation AdS (HV-AdS) geometry. We find
that the linear-T resistivity and quadratic-T inverse Hall
angle can be simultaneously achieved in HV-AdS models, providing a
hint to construct holographic models confronting with the
experimental data of strange metal in future.

\end{abstract} \maketitle

\section{Introduction}

The strange metal phase, which emerges in normal states of
high temperature superconductors and heavy fermion compounds near
a quantum critical point, exhibits a number of weird transport
properties that challenge understanding. The most famous
characteristics of the strange metal phase is that its resistivity
$\rho(T)$ varies linearly with temperature, while the inverse of
Hall angle $\theta_{H}^{-1}(T)$ varies quadratically with
temperature
\cite{Hussey:2008,Hussey:2009,Nakayima:2003,Chien:1991}. Some
theoretical attempts have been made to attack this problem by
proposing that there are different scalings between Hall angle and
resistivity \cite{Anderson:1991,Coleman:1996}. However, due to the
strongly correlated nature in these materials, a complete
resolution to this problem is still absent in theory so far.

AdS/CFT correspondence provides a powerful tool to study strongly
correlated systems. In particular, the methods of computing
transport coefficients by holography have been developed in
\cite{Hartnoll:2009,Herzog:2009}. The linear-T resistivity has
been solely reproduced by holography in
\cite{Polchinski:2010,Zaanen:2014}. One mechanism is to consider a
Lifshitz gravitational system with Dirac-Born-Infeld (DBI) action,
in which the resistivity displays linear property with temperature
when the Lifshitz exponent takes a special value of
$z=2$\cite{Polchinski:2010}. An alternative is proposed in
\cite{Zaanen:2014}. They introduced a hydrodynamic state with a
minimal viscosity $\eta\thicksim s$ being weakly coupled to
disorder, in which the viscosity will contribute to the
resistivity such that $\rho\thicksim \eta\thicksim s$. And then by
holography, a well-controlled locally quantum critical state with
$s\thicksim T$ \cite{Gubser:2010} is introduced such that one has
$\rho\thicksim T$, i.e., the linear-T resistivity. More recently,
progress has been made in addressing different scalings between
Hall angle and resistivity by holography
\cite{Pal:2011,Pal:2012,BSKim:2011,BSKim:2012,BSKim:2013,Gouteraux:2013,Gouteraux:2014,Lee:2011,Sachdev:2014,Karch:2014,Hartnoll:2015,Donos:2014}.
Particularly, a scaling analysis suggests that the anomalous
behaviors $\rho\sim T$ as well as $\theta_{H}\sim 1/T^2$ can be
reproduced in the HV geometry with DBI action\cite{Karch:2014}.
But note that the backreaction of gauge field on the spacetime
geometry is ignored in \cite{Karch:2014}. Another progress is from
Ref. \cite{Donos:2014}, in which the resistivity and Hall angle
exhibit different scalings in holographic Q-lattice model.

Inspired by the work in \cite{Donos:2014}, in present paper we
attempt to address this dichotomy between resistivity and the Hall
angle in a simpler holographic framework, i.e., massive gravity.
The first holographic massive gravity model is constructed in
\cite{Vegh:2013} where a finite DC conductivity is observed.
Subsequently, a lot of works have also implemented finite DC
conductivity in this framework
\cite{Tong:Massive,Amoretti:2014v1,Amoretti:2014v2,Davison:2013,JPWu:Massive,Lucas:2014,Massive:MIT}.
Due to the breaking of diffeomorphism  symmetry, the
stress-energy tensor in massive gravity is not conserved, leading
to a similar effect of dissipating momentum as that of holographic
lattice model
\cite{Horowitz:Lattice1,Horowitz:Lattice2,Horowitz:Lattice3,YLing:Lattice,YLing:CDW,Withers:Lattice,Gouteraux:Lattice,
Donos:QLattice1,Donos:QLattice2,Donos:QLattice3,Donos:QLattice4,XHGv1,XHGv2},
which has originally been pointed out in \cite{Tong:2014}. Here,
we shall first derive general analytic expressions for the
resistivity and the Hall angle in holographic massive EMD gravity
in Section \ref{Hall}. These expressions are
applicable for a larger class of scaling geometries. Specifically, we shall
discuss the case of dyonic RN-AdS and HV-AdS geometry in Section
\ref{HV}. We conclude this paper with a brief discussion on
 the scale dimension analysis at low temperature.

\
\\
\textbf{Note added:} While this work was in preparation
Ref.\cite{Amoretti:2015} appeared, which has some overlap with
ours. In addition, we would like to point out that the
thermoelectric conductivities at finite magnetic field in
holographic Q-lattice model were discussed in
\cite{SJS:1502,Blake:1502}.

\section{DC and Hall conductivity in holographic massive EMD gravity}\label{Hall}

In this section, we shall derive  general analytic
expressions for
 DC and Hall conductivity in holographic massive gravity, which can be applicable for a large class
of scaling geometries. For this purpose, we generalize the
holographic massive gravity action in \cite{Vegh:2013} to the
following EMD theory,
\begin{alignat}{1}
S=\frac{1}{2\kappa^2}\int d^4x\sqrt{-g}\big[R-\frac{Z(\phi)}{4}F^2+\frac{1}{2}(\partial\,\phi)^2+V(\phi)
+\beta(\phi)([\mathcal{K}]^2-[\mathcal{K}^2])\big]\,.
\label{action}
\end{alignat}
where $[\mathcal{K}]:=\mathcal{K}^\mu_{~\mu}$ and
$[\mathcal{K}^2]:=(\mathcal{K}^2)^\mu_{~\mu}$ with
$\mathcal{K}^\mu_{~\nu}:={\sqrt{g^{-1}f}}^{~\mu}_{~~\nu}$ and
$(\mathcal{K}^2)_{\mu\nu}:=\mathcal{K}_{\mu\alpha}\mathcal{K}^{\alpha}_{~\nu}$.
$f_{\mu\nu}$ is the reference metric and we are interested in the
special case with $f_{\mu\nu}=diag(0,0,1,1)$, which breaks the
diffeomorphism along two spatial directions. In comparison with
\cite{Vegh:2013,Amoretti:2015}, a scalar field is introduced with
an arbitrary potential $V(\phi)$, a dilaton-like coupling
$Z(\phi)$ as well as a dilaton dependent coupling parameter
$\beta(\phi)$ of massive term, which is necessary to obtain a
dyonic HV-AdS black hole solution in our setup. When $\beta$ is a
constant, the bulk theory is ghost-free because the lapse function
can be a Lagrange multiplier by choosing a special shift vector,
which is demonstrated in \cite{Vegh:2013}. For details, we
refer to Ref.\cite{Vegh:2013} and references therein. When
$\beta(\phi)$ is a function of scalar field $\phi$, we also find
that our theory (\ref{action}) is ghost-free since $\beta(\phi)$
is still independent of the lapse function.

To study the Hall conductivity, we turn on a magnetic field
on the background and consider the following ansatz
\begin{eqnarray}
&&
\label{Background1}
ds^2=-U(r)dt^2+V(r)dr^2+W(r)(dx^2+dy^2)\,,
\
\\
&&
\label{Background2}
A=a(r)dt+Bxdy\,,~~~~~~~~~~\phi=\phi(r)\,,
\end{eqnarray}
where $B$ is a constant magnetic field. We also assume that
$U,V,W>0$ and at the horizon position $r_+$, $U(r_+)=1/V(r_+)=0,
U(r_+)V(r_+)<\infty$. These assumptions are usually true for
general holographic models.

In order to compute the DC and Hall conductivity, we
consider vector fluctuations over the homogeneous and
isotropic background (\ref{Background1}) and (\ref{Background2}).
Due to the presence of magnetic field, these
perturbations will induce electric currents along $x$ and $y$
directions and provide non-zero contributions to the
$t-x, t-y, r-x, r-y$ components of the energy-momentum tensor. It
turns out that a consistent ansatz for perturbations can be
chosen as
\begin{alignat}{1}
&A_x=-E_xt+a_x(r)\,,\qquad A_y=a_y(r)\,,\label{P1}\\
&\delta g_{tx}:=Wh_{tx}(r)\,,\qquad \delta g_{rx}:=Wh_{rx}(r)\,,\label{P2}\\
& \delta g_{tx}:=Wh_{ty}(r)\,,\qquad \delta g_{ry}:=Wh_{ry}(r)\,.\label{P3}
\end{alignat}
Here, following
closely the method outlined in \cite{Donos:2014,Donos:QLattice2},
we turn on a constant electric field $E_x$  to detect the DC and Hall conductivity.

With the use of Maxwell equations, one can define the
conserved charge $Q$ and the conserved currents $J_x,J_y$ as
follows
\begin{alignat}{1}
&Q:=-Z(\phi)\frac{W}{\sqrt{UV}}a^\prime\,,\label{EP}\\
&J_x:=Qh_{tx}-Z(\phi)B\sqrt{\frac{U}{V}}h_{ry}-Z(\phi)\sqrt{\frac{U}{V}}a_x^\prime\,,\\
&J_y:=Qh_{ty}+Z(\phi)B\sqrt{\frac{U}{V}}h_{rx}-Z(\phi)\sqrt{\frac{U}{V}}a_y^\prime\,,
\end{alignat}
where the prime denotes the derivative with respect to $r$. The
conductivities along $x$ and $y$ directions can be expressed as
$\sigma_{xx}=J_x/E_x, \sigma_{xy}=J_y/E_x$, respectively. Because
$J_x, J_y$ are conserved along $r$ direction, it is more
convenient to evaluate them at $r=r_+$. Thus, the conductivities
can be completely determined by the regularity of fluctuation
modes at the horizon, which are
\begin{eqnarray}
&&
a_x^\prime=-\sqrt{\frac{U}{V}}E_x+\mathcal{O}(r_+-r)\,,~~~~~~~
a_y^\prime=\mathcal{O}(r_+-r)\,,
\
\\
&&
h_{rx}(r_+)=\sqrt{\frac{U}{V}}h_{tx}(r_+)\,,~~~~~~~
h_{ry}(r_+)=\sqrt{\frac{U}{V}}h_{ty}(r_+)\,.
\end{eqnarray}
The currents now can be expressed as
\begin{alignat}{1}
&J_x=Qh_{tx}(r_+)-Z(\phi)|_{r_+}Bh_{ty}(r_+)+Z(\phi)|_{r_+}E_x \label{CC1}\,,\\
&J_y=Qh_{ty}(r_+)+Z(\phi)|_{r_+}Bh_{tx}(r_+)\,.\label{CC2}
\end{alignat}
Taking into account the $x-x, r-x, r-y$ components of Einstein
equation and evaluating them at $r=r_+$, we can derive the
relations between the electric perturbation and metric
perturbation as,
\begin{alignat}{1}
&(B^2Z+m)h_{tx}+QBh_{ty}=QE_x\,,\\
&QBh_{tx}-(B^2Z(\phi)+m)h_{ty}=-ZBE_x\,,
\end{alignat}
where all the variables should be understood as taking values at
$r=r_+$ and the variable $m$ is defined as $m:=-2\beta(\phi)W(r)$.
Putting these solutions into  (\ref{CC1}) and (\ref{CC2}), we
obtain the conductivity as
\begin{alignat}{1}
&\sigma_{xx}=\frac{m(B^2Z^2+Q^2+Zm)}{(B^2Z+m)^2+B^2Q^2}\Big|_{r_+}\,,\\
&\sigma_{xy}=\frac{BQ(B^2Z^2+Q^2+2Zm)}{(B^2Z+m)^2+B^2Q^2}\Big|_{r_+}\,.
\end{alignat}
Finally, the Hall angle $\theta_H:=\sigma_{xx}/\sigma_{xy}$ and DC
conductivity $\sigma_{DC}:=\sigma_{xx}(B=0)$ are derived as
\begin{eqnarray}
\label{Hallv1}
&\theta_H=\frac{BQ(B^2Z^2+Q^2+2Zm)}{m(B^2Z^2+Q^2+Zm)}\Big|_{r_+}\simeq\frac{BQ}{m}\Big|_{r_+}\label{HA}\,,\\
\label{DCv1}
&\sigma_{DC}=Z(\phi)\Big|_{r_+}+\frac{Q^2}{m}\Big|_{r_+}\label{DC}\,.
\end{eqnarray}
A similar result has been reported in \cite{Donos:2014}, in which
a Q-lattice is introduced instead of massive term to dissipate the
momentum. In fact, it has been shown by Andrade and Withers
in \cite{Andrade:2013} that the conductivity in massive
gravity is equivalent to that of a linear axion model, which
could be viewed as a special case of Q-lattices.

Furthermore, we can express the DC conductivity $\sigma_{DC}$ in
terms of the thermodynamical quantities and the generic physical
parameters as
\begin{eqnarray}
\label{DCv2}
\sigma_{DC}=\sigma_{ccs}+\frac{Q^2}{\mathcal{E}+\mathcal{P}}\tau_L
\,,
\end{eqnarray}
where $s=4\pi W(r_+)$, $\mathcal{E}$ and $\mathcal{P}$ are the
entropy density, energy density and pressure of the system,
respectively. In above equation, we have defined a lattice
timescale $\tau_L$ \footnote{$\tau_L$ can be interpreted as a
momentum relaxation rate only in the hydrodynamic regime,
namely at slow momentum dissipation \cite{Davison:2015}.}
\begin{eqnarray}
\tau_L^{-1}=-\frac{s\beta}{2\pi(\mathcal{E}+\mathcal{P})}
\,.
\end{eqnarray}
It is interesting enough to notice that the
coefficient function $\beta(\phi)$ in massive term plays the same
role as the lattice parameter \cite{Donos:2014}, which has
also been revealed in \cite{Tong:2014}.
Also, we defined
\begin{eqnarray}
\sigma_{ccs}=Z(\phi)\Big|_{r_+}
\,,
\end{eqnarray}
which is proposed as a charge-conjugation symmetric conductivity
in \cite{Donos:2014} and independent of the lattice timescale
$\tau_L$ \cite{Karch:2007,Donos:1406.4742}. Moreover, the Hall
angle $\theta_H$ can be expressed in terms of the momentum
relaxation timescale as
\begin{eqnarray} \label{Hallv2}
\theta_H=\frac{BQ}{\mathcal{E}+\mathcal{P}}\tau_L \,.
\end{eqnarray}

So far, we have successfully implemented the dichotomy between DC
resistivity and the Hall angle in the framework of holographic
massive gravity, which provides a new mechanism to realize the
anomalous transport behaviors of strange metal. Since the
functions $Z(\phi)$, $\beta(\phi)$ in Eqs. (\ref{Hallv1}) and
(\ref{DCv1})(or $\sigma_{ccs}$, $\tau_L$ in Eqs. (\ref{DCv2}) and
(\ref{Hallv2})) are quite general and adjustable, following the
ideas proposed in
\cite{Kiritsis:1005.4690,Kiritsis:1107.2116,Kiritsis:1212.2625},
we can build an effective holographic model confronting with the
experimental data. Specifically, we will investigate the
temperature dependence of conductivity and Hall angle over a
dyonic RN-AdS black hole and dyonic HV-AdS black hole in next
sections. In this circumstance, one needs to fix either the
chemical potential or the charge density of the system. From a
phenomenological point of view we intend to perform the analysis
with a fixed charge density since most of the experimental setup
for cuprates are at constant charge density, for instance, as
described in \cite{Ando:2001}.

\section{DC and Hall conductivity in some specific holographic models}\label{HV}

In this section, we shall consider two explicit massive EMD models
in RN-AdS and HV-AdS background, respectively. The corresponding
temperature dependence of DC conductivity and Hall angle are
obtained via the general expressions above. Especially, we find
that the linear-T resistivity and quadratic-T inverse Hall angle
can be achieved simultaneously in a HV model with $z=6/5$ and
$\theta=8/5$.

\subsection{DC and Hall conductivity in dyonic RN-AdS geometry}\label{HallAdS}

For the asymptotic AdS case, we can choose
\begin{alignat}{1}
ds^2=\frac{1}{r^2}\left(-f(r)dt^2+\frac{dr^2}{f(r)}+dx^2+dy^2\right)\,.
\end{alignat}
A background solution for the action with $V(\phi)=6, Z(\phi)=1,
\beta(\phi):=\beta=const.$ in Eq.(\ref{action}) is
\begin{alignat}{1}
&f(r)=1+\beta r^2-Mr^3+\frac{Q^2+B^2}{4}r^4\,,\\
&a(r)=\mu-Qr\,,
\end{alignat}
where the integral constants $\mu,~Q$ can be interpreted as the
chemical potential and charge density respectively, while the
mass $M$ is determined by $f(r_+)=0$. The Hawking temperature is
\begin{alignat}{1}
T=\frac{3}{4\pi r_+}(1+\frac{\beta}{3}r_+^2-\frac{Q^2+B^2}{12}r_+^4)\,,\label{HT1}
\end{alignat}
A straightforward usage of expressions (\ref{Hallv1}) and (\ref{DCv1}) leads to
\begin{alignat}{1}
&\theta_H\simeq-\frac{BQ}{2\beta}r_+^2\,,\\
&\sigma_{DC}=1-\frac{Q^2}{2\beta}r_+^2\,.
\end{alignat}
We are interested in high temperature region, in which we
have the relation $r_+\sim 1/T$. Therefore, when momentum
relaxation is weak (i.e
$\sigma_{DC}\simeq-\frac{Q^2}{2\beta}r_+^2$), the resistivity and
inverse Hall angle would have the similar behavior of temperature
dependence. While when momentum relaxation is strong, the
conductivity $\sigma_{DC}\simeq1$ is independent of the
temperature. So, for both cases, it is hardly possible to
reproduce the linear-T resistivity and quadratic-T inverse Hall
angle simultaneously.

\subsection{DC and Hall conductivity in dyonic HV-AdS geometry}\label{HVRN}

Now, we turn to consider the DC and Hall conductivity in dyonic
HV-AdS geometry including massive term. For this purpose, we
choose
\begin{alignat}{1}
ds^2=r^\theta\left(-\frac{f(r)dt^2}{r^{2z}}+\frac{dr^2}{r^2f(r)}+r^{-2}(dx^2+dy^2)\right)\,,\label{HVa1}
\end{alignat}
where $\theta$ and $z$ are the HV exponent and Lifshitz dynamical
exponent, respectively. The setup of the electromagnetic
field and scalar field is still given by Eq.(\ref{Background2}). In order to obtain a dyonic HV-AdS
solution in an analytical manner, we parameterize the coupling
functions and the potential in the action as
\begin{alignat}{1}
Z(\phi)=Z_0 e^{\lambda\phi}\,,~~~~~
\beta(\phi)=\beta_0e^{\sigma\phi}\,,~~~~~
V(\phi)=V_1 e^{\gamma_1\phi}+V_2e^{\gamma_2\phi}\,.\label{R}
\end{alignat}
An asymptotic HV-AdS black hole solution exists only when the following relations are satisfied
\begin{alignat}{1}
&\lambda=(\theta-2z+2)/\alpha\,,\label{HVa2}\\
&\sigma=-2/\alpha\,,\\
&\gamma_1=-\theta/\alpha\,,\\
&\gamma_2=(-\theta-2z+6)/\alpha\,,\\
&\beta_0=(z-1)(\theta-z-2)\,,\label{MS}\\
&V_1=(\theta-2z)(\theta-z-2)\,,\\
&V_2=\frac{\theta-2z+2}{4(-z+2)}B^2Z_0\,.\label{HVa3}
\end{alignat}
Therefore, the HV model considered here is the action
in Eq.(\ref{action}) with the parameters satisfying
relations (\ref{HVa2}-\ref{HVa3}). Furthermore, we have the
following analytic dyonical HV-AdS black hole solution
\footnote{A detailed derivation of the dyonical HV-AdS black
hole solution is given in the Appendix.}
\begin{alignat}{1}
&e^\phi=\,r^\alpha\,,\qquad \alpha:=\sqrt{(2-\theta)(\theta-2z+2)}\,,\label{Bgsl1}\\
&a(r)=\mu-\frac{Q}{Z_0(z-\theta)}r^{z-\theta}\,,\\
&f(r)=1-Mr^{-\theta+z+2}+\frac{Q^2r^{-2\theta+2z+2}}{2Z_0(\theta-2)(\theta-z)}
+\frac{B^2Z_0r^{-2z+6}}{4(2-z)(\theta-3z+4)}\label{Bgsl2}\,,
\end{alignat}
The Hawking temperature is given by
\begin{alignat}{1}
T=\frac{-\theta+z+2}{4\pi r_+^z}\left(1+\frac{(z-\theta)Q^2r_+^{-2\theta+2z+2}}{2(\theta-z-2)(\theta-2)(\theta-z)}
+\frac{(\theta-3z+4)B^2r_+^{-2z+6}}{4(\theta-z-2)(2-z)(\theta-3z+4)}\right)\,.\label{HT2}
\end{alignat}

Before proceeding, we present several remarks on the dyonic
HV-AdS black hole solution obtained here.
Firstly, we emphasize that
for a given bulk action (Eq.(\ref{action})),
it is only possible to construct the above dyonic solution (Eq.(\ref{Bgsl1})-(\ref{Bgsl2})) for one particular value of the
magnetic field $B$ satisfying Eq. (\ref{HVa3}) (but at generic $T$ and $Q$).
Secondly, we calculate
the specific heat of this black hole and find that only for small
$B$ as well as $\theta<2$, the specific heat is positive. It
indicates that the black hole is thermodynamically stable
only for weak magnetic field and  $\theta<2$. Thirdly, it is easy
to check that this dyonic HV-AdS black hole shares the same near
horizon geometry with dyonic RN-AdS black hole, i.e., $AdS_2\times
\mathbb{R}^2$, but the asymptotic geometry (UV) is hyperscaling
violating characterized by the Lifshitz exponent $z$ and the HV
exponent $\theta$. Note that a different solution for
charged HV-AdS black hole with IR being $AdS_2\times \mathbb{R}^2$
has previously been obtained in \cite{Alishahiha:1209},
and other relevant investigations on EDM models can be found in
\cite{Salvio:2012,Salvio:2013,Dibakar:2014,Dibakar:2015}. In
addition, from Eq.(\ref{MS}) one finds that when $z=1$ and
$\theta=0$, then $\beta(\phi)=0$ and the model goes back to that
without momentum dissipation.

Using Eqs. (\ref{Hallv1}) and (\ref{DCv1}) in section \ref{Hall},
the Hall angle and DC conductivity for the HV model can be
expressed as
\begin{alignat}{1}
&\theta_H\simeq\frac{BQr_+^{4-\theta}}{2(z-1)(2+z-\theta)}\,,\label{HVHA}\\
&\sigma_{DC}=Z_0r_+^{\theta-2z+2}+\frac{Q^2r_+^{4-\theta}}{2(z-1)(2+z-\theta)}\,.\label{HVDC}
\end{alignat}
Obviously, when $z=1$, the DC conductivity becomes infinity due to the absence of momentum dissipation.

Now, we consider the case when the charge-conjugation
symmetric conductivity $\sigma_{ccs}$, i.e., the first term in
(\ref{HVDC}), is dominant over the second one. Using the relation
$r_+\sim T^{-1/z}$ in HV geometry,
we have
\begin{alignat}{1}
\theta_H^{-1}\sim T^{\frac{4-\theta}{z}}\,,~~~~~~~~~~~~~~\rho_{DC}=\sigma_{DC}^{-1}\sim T^{\frac{\theta-2z+2}{z}}
\,.
\end{alignat}
Therefore, the linear-T resistivity and quadratic-T inverse Hall
angle can be simultaneously achieved by setting $z=6/5$ and
$\theta=8/5$ in our holographic massive EMD model with a specific magnetic field $B$.

Finally, we shall provide a scale analysis to ensure that the
results satisfy the scaling law of HV background. Following the
method proposed in \cite{Karch:2014,Hartnoll:2015}, we
introduce the scale dimension of the coordinates and the electric
scalar potential as follows,
\begin{alignat}{1}
[x]=-1\,,\qquad [t]=-z\,,\qquad [ds^2]=-\theta\,,\qquad [A_t]=z-\Phi\,.
\end{alignat}
The scaling of space and time are assigned in terms of the
dispersion relation $\omega\sim k^z$
\cite{Polchinski:2010,Karch:2014,Hartnoll:2015}. The HV exponent
$\theta$ characterizes the scaling of the entropy density.
In addition, an anomalous scaling $\Phi$ is also
introduced for the uniform scaling transformation of the action.
For the detailed discussion, please refer to
\cite{Karch:2014,Hartnoll:2015}. From above equations, one can
derive the scale dimension of other electric variables as
\begin{alignat}{1}
&[E]=1+z-\Phi\,,\qquad [B]=2-\Phi\,,\\
&[Q]=2-\theta+\Phi\,,\qquad [J]=z-\theta+\Phi+1\,.
\end{alignat}
Then, by $\sigma=J/E$, the scale dimension of the conductivity is
\begin{alignat}{1}
\label{sigmaSD}
[\sigma]=-\theta+2\Phi\,.
\end{alignat}
The Hall angle is dimensionless because it is the ratio of two
conductivities $\sigma_{xx},~\sigma_{xy}$. Note that our results
are consistent with those in \cite{Karch:2014,Hartnoll:2015}
with $d=2$.

Now, we check the scale behavior of the DC conductivity
(\ref{HVDC}) and the Hall angle (\ref{HVHA}). It is easy to check
the second term in Eq. (\ref{HVDC}) has the same scale dimension
as that of the conductivity (\ref{sigmaSD}). For the first term in
Eq. (\ref{HVDC}), to have the same scale dimension as that of the
conductivity (\ref{sigmaSD}), we require
\begin{alignat}{1}
[Z(\phi)]=-\theta+2\Phi\,\quad or \quad [Z_0]=-2z+2+2\Phi\,,
\end{alignat}
which is consistent with the uniform scaling transformation condition
\begin{alignat}{1}
\theta=[R]=[(\partial\phi)^2]=[\beta(\phi)r^{-\theta+2}]=[V(\phi)]=[Z(\phi)F^2]\,.
\end{alignat}
Also, it is direct to check that the Hall angle (\ref{HVHA}) we
obtained here is dimensionless.

\section{Conclusions and discussions}

In this paper we have presented a mechanism to implement the
dichotomy between the DC resistivity and the Hall angle in
massive EMD gravity theory, where the diffeomorphism symmetry is
broken along spatial directions and the momentum of the system has
dissipation. Following closely the method performed in
\cite{Donos:QLattice2}, we have derived general analytic
expressions for the DC and Hall conductivity which can be
applicable for a large class of holographic massive models.
Because both gauge field coupling $Z(\phi)$ and massive coupling
$\beta(\phi)$ are completely free and undetermined, this mechanism
can provide a viable road toward an effective holographic field
theory confronting with the experimental data.

As examples, we have presented a detailed analysis on the scaling
behavior of the DC resistivity and Hall angle in dyonic RN-AdS
black hole and dyonic HV-AdS black hole, respectively. The
reproduction of both linear-T resistivity and quadratic-T inverse
Hall angle in dyonic RN-AdS geometry is still suspensive. However,
some surprise occurs in the case of the dyonic HV-AdS geometry
including massive gravity term, in which the linear-T resistivity
and quadratic-T inverse Hall angle can be simultaneously obtained
for $z=6/5$ and $\theta=8/5$ at large $\sigma_{ccs}$.
However, we would like to point out that these scaling behaviors hold only in a specific dyonic HV-AdS geometry with one value of $B$.
In the future, we expect to search for a general dyonic HV-AdS geometry with arbitrary $B$, in which we can achieve simultaneously
these scaling behaviors of strange metal.

We also remark that in this paper a relation $r_+\sim 1/T$ has
been applied as in most previous literature \cite{Karch:2014,Hartnoll:2015,Horowitz:2014}. It is a good
approximation at high temperature when the field parameters $Q$
and $\beta$ are much smaller than $T$ such that the terms
containing $\beta$ and $Q$ in Hawking temperature (\ref{HT1}) can
be ignored. However, at low temperature we must cautiously realize
that this relation may not hold anymore. In this circumstance, one could consider the Taylor expansion of thermal
observables in powers of the scale invariant quantity such as
$T/\sqrt{Q}$ to obtain the behavior of temperature dependence \cite{Andrade:2013}.

\section*{Acknowledgements}
We are grateful to the anonymous referees for valuable suggestions and comments,
which are important in improving our work.
We also thank A. Amoretti and A. Karch for helpful discussions.
This work is supported by the Natural
Science Foundation of China under Grant Nos.11275208, 11305018 and
11178002. Y.L. also acknowledges the support from Jiangxi young
scientists (JingGang Star) program and 555 talent project of
Jiangxi Province. J. P. Wu is also supported by Program for
Liaoning Excellent Talents in University (No. LJQ2014123).

\begin{appendix}
\section{The dyonic HV-AdS solution of massive EMD gravity }\label{Appendix1}

In this appendix, we present a detailed derivation about the
dyonic HV-AdS solution in \ref{HVRN}. Firstly, we rewrite the
action in Eq.(\ref{action}) as follows
\begin{eqnarray}
S=\int d^4x \sqrt{-g}(R-\frac{Z(\phi)}{4}F^2+\frac{1}{2}(\partial\phi)^2+V(\phi)+2\beta(\phi)\sqrt{g^{xx}g^{yy}})
\label{ActionAv1}
\,.
\end{eqnarray}
To obtain a dyonic HV-AdS solution, we parameterize $Z(\phi)$,
$V(\phi)$ and $\beta(\phi)$ by exponentials in terms of $(\lambda,
\sigma, \gamma_1, \gamma_2, Z_0, \beta_0, V_1, V_2)$ (see Eq.
(\ref{R})). At the same time, we assume a dyonic HV-AdS ansatz
\begin{eqnarray}
\label{Av1}
&&ds^2=r^\theta\left(\frac{-f(r)}{r^{2z}}dt^2+\frac{dr^2}{f(r)r^2}+r^{-2}(dx^2+dy^2)\right)\,,\\
\label{Av2}
&&A=a(r)dt+Bxdy\,,~~~~~~\phi=\phi(r)
\,,
\end{eqnarray}
where we have parameterized the metric by the Lifshitz dynamical
exponent $z$ as well as the HV exponent $\theta$ and introduced a
constant magnetic field $B$.

Applying the variational approach to the action in
(\ref{ActionAv1}), one can derive the Einstein equations as
\begin{eqnarray}
R^\mu_{~\nu}=M^\mu_{~\nu}-S^\mu_{~\nu}
\,,
\end{eqnarray}
with
\begin{eqnarray}
&&M_{\mu\nu}\equiv\frac{Z}{2}(F_{\mu\alpha}F_\nu^{~\alpha}-\frac{1}{4}g_{\mu\nu}F^2)\,,\\
&&S_{\mu\nu}\equiv\frac{\delta\mathcal{L}_m}{\delta g_{\mu\nu}}-\frac{1}{2}g_{\mu\nu}g^{\alpha\beta}\frac{\delta\mathcal{L}_m}{\delta g_{\alpha\beta}}+\frac{1}{2}g_{\mu\nu}\mathcal{L}_m\,,\\
&&\mathcal{L}_m\equiv\frac{1}{2}(\partial\phi)^2+V(\phi)+2\beta(\phi)\sqrt{g^{xx}g^{yy}}
\,.
\end{eqnarray}
Using the ansatz (\ref{Av1}) and (\ref{Av2}), $M^\mu_{~\nu}$ and
$S^\mu_{~\nu}$ can be expressed as
\begin{eqnarray}
&&M^\mu_{~\nu}=\frac{g^{xx}g^{yy}(Q^2+B^2Z^2)}{4Z}diag\{-1,-1,1,1\}\,,\\
&&S^\mu_{~\nu}=diag\{\frac{V}{2},\frac{1}{2}g^{rr}(\phi^\prime)^2+\frac{V}{2},\beta\sqrt{g^{xx}g^{yy}}+\frac{V}{2},\beta\sqrt{g^{xx}g^{yy}}+\frac{V}{2}\}
\,,
\end{eqnarray}
where the conserved charge $Q\equiv-Z\sqrt{-g}g^{tt}g^{rr}a^\prime$ has been defined using the Maxwell equations.
Thus, we can express the Einstein equations as
\begin{eqnarray}
&&R^t_{~t}=-\frac{g^{xx}g^{yy}(Q^2+B^2Z^2)}{4Z}-\frac{V}{2}\,,\label{Et}\\
&&R^r_{~r}=-\frac{g^{xx}g^{yy}(Q^2+B^2Z^2)}{4Z}-\frac{1}{2}g^{rr}(\phi^\prime)^2-\frac{V}{2}\,,\label{Er}\\
&&R^x_{~x}=\frac{g^{xx}g^{yy}(Q^2+B^2Z^2)}{4Z}-\beta\sqrt{g^{xx}g^{yy}}-\frac{V}{2}\,,\label{Ex}\\
&&R^y_{~y}=\frac{g^{xx}g^{yy}(Q^2+B^2Z^2)}{4Z}-\beta\sqrt{g^{xx}g^{yy}}-\frac{V}{2}
\,,
\end{eqnarray}
as well as the scalar field equation
\begin{eqnarray}
\triangle\phi=\frac{\dot{Z}g^{xx}g^{yy}(Q^2-B^2Z^2)}{2Z^2}+2\dot{\beta}\sqrt{g^{xx}g^{yy}}+\dot{V}
\,,\label{seq}
\end{eqnarray}
where $\triangle$ is the Laplace operator and dot denotes the
derivative with respect to $\phi$. To solve them, we will use the
following two important relations
\begin{eqnarray}
&&R^r_{~r}-R^t_{~t}=\frac{f}{2r^\theta}(\theta-2)(\theta-2z+2)\,,\label{R1}\\
&&r(R^x_{~x})^\prime+2(\theta-z)R^x_{~x}=(\theta-2)R^t_{~t}\label{R2}
\,.
\end{eqnarray}
From the first one and Eqs.(\ref{Et}) and (\ref{Er}) , we can obtain the solution of $\phi$
\begin{eqnarray}
\phi^\prime=\alpha/r\,,~~~~~~\alpha^2\equiv(2-\theta)(\theta-2z+2)
\,.\label{ssl}
\end{eqnarray}
Since the second relation (\ref{R2}) together with Eq.(\ref{Ex})
implies Eq.(\ref{Et}), we leave only two equations
\begin{eqnarray}
&&\frac{2-\theta}{2r^\theta}((\theta-z-2)f+rf^\prime)=\frac{(Q^2+B^2Z^2)}{4Zr^{2(\theta-2)}}-\beta r^{2-\theta}-\frac{V}{2}\,,\label{im1}\\
&&\frac{rZ^\prime (Q^2-B^2Z^2)}{4Z^2r^{2(\theta-2)}}+\beta^\prime r^{3-\theta}+\frac{V^\prime r}{2}=(\theta-2z+2)(\frac{Q^2+B^2Z^2}{4Zr^{2(\theta-2)}}-\beta r^{2-\theta}-\frac{V}{2})
\,,\label{im2}
\end{eqnarray}
where the first one is from the x-x component of the Einstein
equation (\ref{Ex}) and the second one comes from the scalar field
equation (\ref{seq}).

Now, substituting the solution of $\phi$ (Eq.(\ref{ssl})) into
the coupling functions and the potential, one has
\begin{alignat}{1}
Z(\phi)=Z_0 r^\lambda\,,~~~~~
\beta(\phi)=\beta_0r^{\sigma}\,,~~~~~
V(\phi)=V_1 r^{\gamma_1}+V_2r^{\gamma_2}\,,
\end{alignat}
Then, Eq.(\ref{im1}) gives
\begin{alignat}{1}
f(r)=&-\frac{Q^2r^{4-\theta-\lambda }}{2Z_0(\theta-2)(2-\lambda -z)}
-\frac{B^2Z_0r^{4-\theta+\lambda }}{2(\theta-2)(2+\lambda -z)}\nonumber\\
&+\frac{2\beta_0r^{2+\sigma }}{(\theta-2)(\theta+\sigma -z)}
+\frac{V_{1}r^{\theta+\gamma_1 }}{(\theta-2)(2\theta+\gamma_1 -z-2)}\nonumber\\
&+\frac{V_{2}r^{\theta+\gamma_2 }}{(\theta-2)(2\theta+\gamma_2 -z-2)}
-Mr^{-\theta+z+2}\,,
\end{alignat}
where $M$ is an integral constant. Eq.(\ref{im2}) becomes a
polynomial equation of $r$
\begin{alignat}{1}
(\theta-2z+2-\lambda)\frac{Q^2r^{4-\theta-\lambda}}{2Z_0}+(\theta-2z+2+\lambda)\frac{B^2Z_0r^{4-\theta+\lambda}}{2}
-(\theta-2z+2+\sigma)2\beta_0r^{2+\sigma}\nonumber \\
-(\theta-2z+2+\gamma_1)V_1r^{\theta+\gamma_1}-(\theta-2z+2+\gamma_2)V_2r^{\theta+\gamma_2}=0\,.\label{im3}
\end{alignat}
To obtain an asymptotic HV-AdS solution to above two
equations, one finds that the parameters should satisfy the
following relations
\begin{alignat}{1}
&\lambda=\theta-2z+2,~~~~\sigma=-2,~~~~\gamma_1=-\theta,~~~~\gamma_2=-\theta-2z+6\nonumber,\\
&\beta_0=(z-1)(\theta-z-2)~~~~V_1=(\theta-2z)(\theta-z-2)~~~~V_2=\frac{(\theta-2z+2)B^2Z_0}{-4z+8}\,.
\label{Arelation}
\end{alignat}
 Then, we have
\begin{alignat}{1}
f(r)=1-Mr^{-\theta+z+2}+\frac{Q^2r^{-2\theta+2z+2}}{2Z_0(\theta-2)(\theta-z)}
+\frac{B^2Z_0r^{-2z+6}}{4(2-z)(\theta-3z+4)}\,.
\end{alignat}
In addition, the solution of electric potential $a(r)$ can be
obtained from Eq.(\ref{EP}), given by
\begin{alignat}{1}
a(r)=\mu-\frac{Q}{Z_0(z-\theta)}r^{z-\theta}\,.
\end{alignat}

So far, we have obtained an analytical dyonic HV-AdS black hole solution.
In order to support an analytical dyonic HV-AdS black hole solution from the
action (Eq.(\ref{ActionAv1})), the parameters $(\lambda, \sigma,
\gamma_1, \gamma_2, \beta_0, V_1, V_2)$ in Eq. (\ref{R}) must be
expressed in terms of the parameters $z$, $\theta$ and $B$ in the
ansatz (Eqs.(\ref{Av1}) and (\ref{Av2})), i.e., satisfying
Eqs.(\ref{Arelation}).
This method has been applied to obtain the
analytical Lifshitz-AdS and HV-AdS in
\cite{Lifshitz:1105,Alishahiha:1209}, respectively.

\end{appendix}

\end{document}